\documentclass[12pt]{iopart}
\pdfoutput=1 
\usepackage[english]{babel}
\usepackage{graphicx}
\usepackage{subfigure} 
\usepackage{iopams} 
\usepackage{float}

\usepackage{color}
\definecolor{darkgreen}{rgb}{0,0.6,0}
\definecolor{darkblue}{rgb}{0,0,0.6}
\definecolor{darkred}{rgb}{0.6,0,0}
\definecolor{darkpurple}{rgb}{0.5,0,0.5}
\usepackage[colorlinks=true,urlcolor=darkblue,citecolor=darkblue,linkcolor=darkred,hyperfootnotes=false]{hyperref}

\usepackage{etoolbox}

\makeatletter
\def\@mkboth#1#2{}
\newlength\appendixwidth
\preto\appendix{\addtocontents{toc}{\protect\patchl@section}}
\newcommand{\patchl@section}{%
  \settowidth{\appendixwidth}{\textbf{Appendix }}%
  \addtolength{\appendixwidth}{1.5em}%
  \patchcmd{\l@section}{1.5em}{\appendixwidth}{}{\ddt}%
}
\makeatother

\begin{document}



\title[\small Finite-Time and Finite-$N_c$ Scalings in the Large-$L$ Limit]{Breakdown of the Finite-Time and -Population Scalings of the Large Deviation Function in the Large-Size Limit of a Contact Process}

\author{Esteban Guevara Hidalgo}

\address{Institut Jacques Monod, CNRS UMR 7592, Universit\'{e} Paris Diderot, Sorbonne Paris Cit\'{e}, F-750205, Paris, France}
\address{Laboratoire de Probabilit\'{e}s, Statistique et Mod\'{e}lisation, CNRS UMR 8001, Universit\'{e} Paris Diderot, Sorbonne Paris Cit\'{e}, 75013 Paris, France}
 
\ead{esteban\_guevarah@hotmail.com} 
\vspace{10pt}

\begin{abstract}
In a recent study~\cite{partI,partII}, the finite-time ($t$) and -population size ($N_c$) scalings in the evaluation of a large deviation function (LDF) estimator were analyzed by means of the cloning algorithm.
These scalings provide valuable information about the convergence of the LDF estimator in the infinite-$t$ and infinite-$N_c$ limits. 
For the cases analyzed in that study, the scalings of the systematic errors of the estimator were found to behave as $t^{-1}$ and $N_c^{-1}$ in the large-$t$ and large-$N_c$ asymptotics. Moreover, it was shown how 
this convergence speed can be used in order to extract an asymptotic limit which resulted to render a better LDF estimation in comparison to the standard estimator. However, the validity of these scaling laws and thus, the convergence of the estimator was proved only in systems for which  the number of sites $L$ (where the dynamics occurs) was small. In this paper, the analysis is extended to a wider range of system sizes $L$. We show how the introduction of the exponents $\gamma_{t}$ and $\gamma_{N_c}$ allows to characterize the behavior of the LDF estimator for any system size. From these generalized $t^{-\gamma_{t}}$- and $N_{c}^{-\gamma_{N_c}}$- scalings, we verify that in the large-$L$ limit the $t^{-1}$- and $N_c^{-1}$-scalings are no longer valid. Moreover, 
as the convergence of the estimator relies on the positivity of these exponents, we show how for some cases $\gamma_{N_c}$ can be negative implying that the estimation provided by the cloning algorithm is no longer reliable.

\vspace{2pc}


\end{abstract}

\newpage 
\tableofcontents

%

%
%
%
%
%
\section{Introduction}

In order to study the properties of rare events and rare trajectories in stochastic dynamics, a large variety of methods have been developed~\cite{touchette_large_2009,giardina_simulating_2011, bucklew_introduction_2013}.
The numerical approaches range from importance sampling~\cite{kahn1951estimation}, 
to ``go with the winner'' algorithms~\cite{aldous1994go,GRASSBERGER200264},
adaptive multilevel splitting~\cite{cerou_adaptive_2007} 
and transition path sampling~\cite{bolhuis_transition_2002}.
Through this paper we will give particular attention to the population dynamics algorithms~\cite{giardina_direct_2006, lecomte_numerical_2007, tailleur_simulation_2009}.
Under this approach, 
the study of rare trajectories in a system is done by exponentially biasing their probability. The resulting modified dynamics consists in the coupled evolution of a large number of copies (or clones) $N_c$ of the original process supplemented with a selection rule according to which a copy of the system is multiplied if it is rare or killed, if it is not.
%

The distribution of the class of rare trajectories in the original dynamics is related with the exponential growth of the population of clones of the system and an estimator for LDF can be obtained from its growth rate (the cumulative generating function CGF).
The numerical determination of this estimator is systematized in a method known as the cloning algorithm (which can be performed in a number of ways~\cite{giardina_direct_2006, lecomte_numerical_2007, tailleur_simulation_2009, giardina_simulating_2011,hidalgo_discreteness_2016,partI, partII}). 
However, this method introduces two additional parameters into consideration:
the population size
%
%
$N_c$ and the simulation time $t$. Both of which affect considerably the accuracy of the CGF estimation which is expected to be high in the infinite-$t$ and infinite-$N_c$ limit. 
Given that this is not achievable in practice, what is generally done is to choose 
these parameters
large enough such that the average estimator (over several realizations of the algorithm) does not depend on them.


The finite-$t$ and finite-$N_c$ scalings in the evaluation of the CGF
provide useful information about the convergence of this estimator in the infinite-$t$ and infinite-$N_c$ limits. They were analyzed recently following two different approaches: an analytical one, in Ref.~\cite{partI}, using a discrete-time version of the population dynamics algorithm~\cite{giardina_direct_2006}, and a numerical one, in Ref.~\cite{partII}, using a continuous-time version~\cite{lecomte_numerical_2007,tailleur_simulation_2009}. 
In both cases, the systematic errors of these scalings were found to behave as $1/t$ and $1/N_c$ in the large-$t$ and large-$N_c$ asymptotics respectively. 
Moreover, it was shown how these scaling properties can be used in order to improve the CGF estimation (as shown in Ref.~\cite{partII}). 
This is done considering that the asymptotic behavior of the estimator in the $t \to \infty$ and $N_c \to \infty$ limits may be interpolated from the data obtained from simulations at finite (and relative small) simulation time and number of clones. 
The improvement in the CGF estimation was illustrated on a simple two-states annihilation-creation dynamics (in one site) and on a more complex system, a contact process~\cite{CP, GRASSBERGER1979373, liggett2012interacting} (with $L=6$ sites). However, the validity of these scalings and thus, the convergence of the estimator
as the number of sites $L$ increases was left as pending. This is precisely the purpose of this paper where we complement the results presented in Ref.~\cite{partII} by extending the analysis of the finite scalings of the CGF estimator to a large-$L$ contact process. 

The paper is organized as follows. 
In Sec.~\ref{sec:LD} we introduce the method used in order to estimate large deviations of additive observables. 
The finite-time and finite-$N_c$ scalings of the CGF are summarized in Sec.~\ref{subsec: tnScaling}, 
%
and generalized to large-$L$ systems in Sec.~\ref{subsec: tnScalingL}. 
We make use of these results in Sec.~\ref{sec: CPL100} where we check their validity 
(Sec.~\ref{sec: CGFL100}), their behavior 
(Sec.~\ref{sec: gamma_tn}), as well as the convergence of the CGF estimator 
(Sec.~\ref{sec: SML100}) for a contact process with $L=100$ sites.
This analysis is generalized in Sec.~\ref{sec: planeSL} where we characterize the finite CGF scalings in the plane $s-L$. Before presenting our conclusions in Sec.~\ref{sec:conclusion}, we discuss about the effects of the dynamical phase transition 
in Sec.~\ref{sec:DPTcp}. The description of the models and methods used throughout the paper can be found at the~\ref{sec:App}.

\section{Biased Markov Dynamics and the Cloning Algorithm}
\label{sec:LD}
In order to analyze the large deviations of the activity in the contact process, we will make use of the continuous-time version of the cloning algorithm~\cite{lecomte_numerical_2007, tailleur_simulation_2009, giardina_simulating_2011}. 
This approach allows to obtain an estimator of LDF (the cumulative generating function) from the exponential growth (or decay) rate of a set of copies of the system which evolves following a ``$s$-modified dynamics''. This procedure is summarized below. For a description of the contact process refer to~\ref{sec:CP}.

\subsection{Additive Observables and their Large Deviations}
We consider a general Markov dynamics which evolves continuously in time. The system jumps from configuration $C$ to $C'$ with transition rates $W(C \rightarrow C')$. The probability $P(C,t)$ to find the system at time $t$ in configuration $C$ verifies the master equation
\begin{equation}\label{eq:1}
\partial_{t}P(C,t)={\sum\limits_{C' }^{ }} W(C'\rightarrow C)P(C',t) - r(C)P(C,t),
\end{equation}
where $r(C) = {\sum\limits_{C' }}  W(C \rightarrow C')$ is the escape rate from configuration $C$. 

A trajectory of $K$ configurations jumps, $(C_0,\ldots,C_K)$, can be characterized by some additive observable $\mathcal O$ (extensive in time) which is defined as 
\begin{equation}
\label{eq:obs}
\mathcal O =
\sum_{k=0}^{K-1} a(C_k,C_{k+1}),
\end{equation}
where $a$ 
describe elementary increments. The joint distribution $P(C,\mathcal O,t)$ describes the probability of finding the system in the configuration $C$, with a value of the observable $\mathcal O$, and at time $t$. 
On the other hand, the probability of observing an atypical value 
$\hat{o} = \mathcal O/t$ of observable $\mathcal O$ after a large time scales as 
\begin{equation}
\label{eq:ldpO}
P(\hat{o}= \mathcal O/t,t) \sim e^{t\pi(\hat{o})},
\end{equation}
in the large time asymptotics. Equation~(\ref{eq:ldpO}) is known as the large deviation principle for observable $\mathcal O$~\cite{touchette_large_2009}. The problem, then reduces to the determination of the rate function $\pi(\hat{o})$ which is known as large deviation function. For practical purposes, it is convenient to consider instead its Legendre transform $\psi(s)$ which is called scaled cumulant generating function (CGF).
The procedure followed in order to analyze the large deviations of these observables~(\ref{eq:obs}), 
consists in biasing the statistical weight of histories of the system by a parameter $s$ (conjugated to $\mathcal O$)~\cite{lecomte_numerical_2007,tailleur_simulation_2009}. A value of $s$ different from zero favors the non-typical values of the observable $\mathcal O$ whose average value has been fixed.
In the $t\to\infty$ limit, the corresponding dynamical partition function scales as
\begin{equation}
\label{eq:Zeta}
Z(s,t)=\langle e^{-s\mathcal O}\rangle\sim e^{t\psi(s)},
\end{equation}
A main feature related with $\psi(s)$ is that its derivatives in $s=0$ allow to recover the large-time limit of the cumulants of $\mathcal O$~\cite{touchette_large_2009}. 


%

\subsection{Continuous-Time Population Dynamics}
Taking into consideration the relation between the dynamical partition function
\begin{equation}
Z(s,t)=\sum_C\hat P(C,s,t),
\end{equation}
and the Laplace transform of the distribution $P(C,\mathcal O,t)$
\begin{equation}
\label{eq:Z}
\hat P(C,s,t) = \int d\mathcal O\: e^{-s\mathcal O} P(C,\mathcal O,t),
\end{equation}
the original dynamics (characterized by a value of the observable $\mathcal O$) can be transformed into a ``$s$-modified'' one. This biased dynamics verifies the time-evolution equation~\cite{garrahan_first-order_2009}
\begin{equation}
 \label{eq:Phat}
 \partial_t  \hat P(C,s,t) = \sum\limits_{C'}^{ } (\mathbb W_s)_{C'C} \hat P(C',s,t) + \delta r_s( C) \hat P(C,s,t),
\end{equation}
where 
\begin{equation}
\label{eq:opWs}
(\mathbb W_s)_{C'C}= W_s(C'\to C) -r_s(C)\delta_{CC'},
\end{equation}
and
\begin{equation}
\label{eq:deltaR}
\delta r_s( C) = r_s(C)-r(C).
\end{equation}
The expression 
\begin{equation}
  \label{eq:Ws}
  W_s(C\to C') = e^{-s a(C,C')} W(C\to C'),
\end{equation}
represent a $s$-modified transition rate, whereas
  \label{eq:rs}
\begin{equation}
  r_s(C) = \sum_{C^{\prime}} W_s(C\to C')
\end{equation}
is the corresponding biased escape rate.
Equation~(\ref{eq:Phat}) can be interpreted as a population dynamics of a large number $N_c$ of copies of the system which evolves 
with transition rates $W_s(C\to C')$ and with a selection mechanism of rates $ \delta r_s( C)$~\cite{giardina_direct_2006,tailleur_probing_2007}. Depending on $ \delta r_s( C)$, a copy of the system is multiplied or killed, so that under this $s$-biased dynamics an atypical class of histories of the original process becomes typical.

A numerical estimator for $\psi(s)$ in Eq.~\ref{eq:Zeta}, that we will denote as $\Psi_{s}^{(N_{c})}$, can be obtained from the exponential growth (or decay) rate of these population of copies of the system evolving with rules mentioned above. The method which systematize the numerical determination of this estimator is know as \textbf{cloning algorithm} which can be performed in a number of ways~\cite{giardina_direct_2006,partI, tailleur_simulation_2009, lecomte_numerical_2007, giardina_simulating_2011,hidalgo_discreteness_2016}. A detailed description of the version 
used through this paper can be found 
%
in the~\ref{sec:AppCA}. By other hand, the quantity $\mathcal O$ whose ``large deviations'' are analyzed is the dynamical activity $K$ which is the number of configuration changes on the time interval $[0,t]$. 

\subsection{CGF Numerical Estimator}
\label{subsec: CGF}
Using the constant-population approach of the continuous-time cloning algorithm on a $s$-biased Markov dynamics, the average over $R$ realizations of the CGF estimator $\Psi_{s}^{(N_{c})}$ for $N_{c}$ clones (or copies of the system) and a final simulation time $T$, is defined as
\begin{equation} \label{eq:PSI1}
\overline { \Psi_{s}^{(N_{c})} }  = \frac{1}{R} \sum\limits_{r = 1}^{R}  \frac{1}{t_{r}^{\mathcal{F}}}\log\prod \limits_{i = 1}^{\mathcal K_{r}} X_{i}^{r},
\end{equation}
where  $\mathcal K$ is the total number of configuration changes in the full population up to time $t_{r}^{\mathcal{F}} \lesssim T$. At each configuration change, the population of clones is increased by a factor $X_{i}^{r}$ where $ r\in\{1,...,R\}$. The time $t_{r}^{\mathcal{F}}$ is the actual final simulation time. For $t_{r}^{\mathcal{F}}>>0$, (as discussed in Ref.~\cite{hidalgo_discreteness_2016}) 
$ \big \vert \overline { \Psi_{s}^{(N_{c})}(T) }  - \overline { \Psi_{s}^{(N_{c})} (t_{r}^{\mathcal{F}}) } \big \vert \approx 0$ and thus, it is possible to set $t_{r}^{\mathcal{F}} \approx T $ in Eq.~(\ref{eq:PSI1}), so that we can make use of the expression
\begin{equation} \label{eq:PSI2}
 \overline { \Psi_{s}^{(N_{c})}(T)}  \simeq \frac{1}{RT} \sum\limits_{r = 1}^{R}  \log\prod \limits_{i = 1}^{\mathcal K_{r}} X_{i}^{r}.
\end{equation}
%
For an extensive discussion of the dependence of this estimator with $R$ refer to the Appendix C of Ref.~\cite{partII}.
It is expected that in the infinite-$t$ and infinite-$N_c$ limits, Eqs.~(\ref{eq:PSI1}) and~(\ref{eq:PSI2}) provide an accurate estimation of the CGF, i.e.,
\begin{equation}
\lim_{N_c \to \infty} \lim_{t \to \infty} \overline{\Psi_{s}^{(N_{c})}(t)} = \psi(s).
\label{eq:limitNcT}
\end{equation}
However, as these limits are not achievable in practice, the best estimation can be obtained considering a large enough simulation time $T$ and number of clones $N_c$. 
The dependence of the estimator with these parameters was studied in Refs.~\cite{partI,partII} and summarized in Sec.~\ref{subsec: tnScaling}.

\section{Finite Scalings of the CGF Estimator}
\label{sec: Scaling}

The approach described in Sect.~\ref{sec:LD} 
was followed in Ref.~\cite{partII} in order to
compute an estimator of the large deviations
%
of the activity for two specific models: a simple one-site annihilation-creation dynamics and a contact process (as described in~\ref{sec:CP}) with $L=6$ sites.
The accuracy of the method can be tested
by comparing the estimator $\overline { \Psi_{s}^{(N_{c})}(T)}$ with the corresponding analytical expression of the CGF $\psi$. However this can be done only for the cases for which $\psi$ can be computed exactly (see Ref.~\cite{touchette_large_2009} for a review). 
%
%
For the cases presented in Ref.~\cite{partII}, $\psi$ can be obtained as the largest eigenvalue of the operator $\hat{\mathbb W}_s$ in Eq.~(\ref{eq:Phat}), $\partial_t \hat{P}(C,s,t)  = \hat{\mathbb W}_s \hat{P}(C,s,t)$.
This	 allowed to present a clear picture of the dependence of the estimator $\overline { \Psi_{s}^{(N_{c})}(T)}$~(\ref{eq:PSI2})
with the time and the number of clones $N_c$ and importantly, its convergence to $\psi$. Moreover, the 
\textbf{scaling behavior} of the CGF estimator was 
consistent
for both models and its speed convergence was used in order to improve its estimation.
%
However, whether this behavior was valid for larger system sizes $L$ or not was left as an open problem.
%
Below, we summarize the finite-time and finite-$N_c$ scalings of the CGF estimator (as presented in Refs.~\cite{partI,partII}) and its generalization to large-$L$ systems.

\subsection{Large-Time and Large-$N_{c}$ Limit}
\label{subsec: tnScaling}

When we analyze the time behavior of the CGF estimator~(\ref{eq:PSI2}) 
for a fixed number of clones $N_c$,
we observe 
this can be well described by a curve $f_{t}^{(N_{c})}$~(\ref{eq:tScal1}) indicating the existence of a $t^{-1}$-convergence to the value $f_{\infty}^{(N_{c})}$. We call this $\mathbf{t^{-1}}$\textbf{-scaling} and it is valid independently if $N_c$ is small or large. 
The curve $f_{t}^{(N_{c})}$ is determined from a fit in time over $\overline{ \Psi_s^{(N_c)}(t)}$ up to (the final simulation) time $T$. From this fit, it is possible to determine the infinite-time limit of the CGF estimator $f_{\infty}^{(N_{c})}=\lim_{t\rightarrow \infty}\overline{ \Psi_{s} ^{(N_{c})}(t)}$ 
\footnote{Additionally, the behavior of the standard CGF estimator $\overline{ \Psi_{s}^{(N_{c})} (T) }$ as a function of the population size $N_c$ is well described by a behavior of the form
\begin{equation*}
g_{N_{c}}^{(T)} = g_{\infty}^{(T)} + \tilde b_{N_{c}}^{(T)}N_{c}^{-1}
\end{equation*}
indicating that $\overline{ \Psi_{s}^{(N_{c})} (T) }$ also converges to its infinite-$N_c$ limit $g_{\infty}^{(T)}=  \lim_{N_c\to \infty}  \overline{ \Psi_{s}^{(N_{c})} (T) }$ with an error proportional to $1/N_c$.}.

When we repeat this procedure for different values of population size $N_c \in \vec{N}_{c}= \{ N_{c}^{(1)},...,N_{c}^{(j)} \}$, extracting in each case the corresponding 
$f_{\infty}^{(N_{c})}$'s, we observe they exhibit $1/N_c$ corrections in $N_c$ ($\mathbf{N_{c}^{-1}}$ \textbf{-scaling}).
In other words, the $f_{\infty}^{(N_c)}$'s satisfy a equation of the form~(\ref{eq:nScal1}) which can be obtained from a fit in $N_{c}$ over the extracted $f_{\infty}^{(N_{c})}$'s. 
Thus, the $t^{-1}$- and $N_{c}^{-1}$-scalings of the CGF estimator are given by
\begin{eqnarray}
\label{eq:tScal1}
f_{t}^{(N_{c})}  = f_{\infty}^{(N_{c})} + b_{t}^{(N_{c})} t^{-1}, \\
\label{eq:nScal1}
f_{\infty}^{(N_c)} = f_{\infty}^{\infty} + b_{\infty}^{(N_c)} N_c^{-1}.
\end{eqnarray}
%
These equations imply that $\overline{ \Psi_s^{(N_c)}(t)}$ converges to its infinite-$t$ and infinite-$N_c$ limit, $f_{\infty}^{\infty} = \lim_{N_{c}\to\infty} f_\infty^{(N_c)}$, proportionally to $1/t$ and $1/N_c$. 
Importantly, this limit can be obtained using a small number of clones and simulation time by making use of the \textbf{scaling method}~\cite{partII}~(see \ref{sec:AppSM}). The results obtained for $f_{\infty}^{\infty}$ rendered a better estimation of $\psi(s)$ than the \textbf{standard CGF estimator} which is obtained from evaluating $\overline{ \Psi_{s}^{(N_{c})}(t) }$ for the largest number of clones $N_c = \max \vec{N_c}$ and for $t = T$.

\subsection{Scalings in the Large-$L$ Limit}
\label{subsec: tnScalingL}
In order to verify whether the 
scalings laws observed in small 
systems are also valid in the large-$L$ limit, 
%
%
we assume that the CGF estimator 
can be described by equations of the form
\begin{eqnarray}
\label{eq:tScal2}
\chi_{t}^{(N_{c})}  \equiv 
& \chi_{\infty}^{(N_{c})} + \kappa_{t}^{(N_{c})} t^{-\gamma_{t}}, \\
\label{eq:nScal2}
\chi_{\infty}^{(N_c)} \equiv 
&\chi_{\infty}^{\infty} + \kappa_{\infty}^{(N_c)} N_c^{-\gamma_{N_{c}}},
\end{eqnarray}
redefining in a more general way the scalings~(\ref{eq:tScal1}) and~(\ref{eq:nScal1}).
We will refer to Eq.~(\ref{eq:tScal2}) as \textbf{$t^{-\gamma_{t}}$-scaling} whereas Eq.~(\ref{eq:nScal2}) as \textbf{$N_{c}^{-\gamma_{N_{c}}}$-scaling}. 
The problem reduces in determining the exponents $\gamma_{t}$ and $\gamma_{N_{c}}$ in order to verify if effectively $\gamma_{t} \approx 1$ and $\gamma_{N_{c}} \approx 1$ and
whether the terms $\chi_{\infty}^{(N_{c})}$ and $\chi_{\infty}^{\infty}$ represent the limits in $t \to \infty$ and $N_{c} \to \infty $ of the CGF estimator.
Thus, a value of the exponent $\gamma_{t} \approx 1$, verifies $\chi_{\infty}^{(N_{c})} \approx f_{\infty}^{(N_{c})} $ and $\gamma_{N_{c}} \approx 1$, verifies $\chi_{\infty}^{\infty} \approx f_{\infty}^{\infty}$.
This is done in Sec.\ref{sec: CPL100} on a contact process with $L=100$ sites. Below we describe the procedure followed in order to obtain these exponents \footnote{\label{foot:Gamma_n} Additionally to Eqs.~(\ref{eq:tScal2}) and~(\ref{eq:nScal2}), the $N_c$-behavior of $\overline{ \Psi_{s}^{(N_{c})} (T) }$ can be described by the equation
\begin{equation}
\chi_{N_{c}}^{(T)} = \chi_{\infty}^{(T)} + \tilde \kappa_{N_{c}}^{(T)}N_{c}^{-\gamma_{N_{c}}^{T}},
\label{eq:nScal3}
\end{equation}
where $\chi_{\infty}^{(T)} = \lim_{N_{c}\to\infty} \overline{ \Psi_{s}^{(N_{c})} (T) }$. Here it is important to remark that both $\chi_{\infty}^{(N_{c})}$ and $\overline{ \Psi_{s}^{(N_{c})} (T) }$ scale in the same way in $N_c$. In other words,
 $\gamma_{N_{c}} \approx \gamma_{N_{c}}^{T}$.}.


\subsubsection{Determination of the Exponents $\gamma_{t}$ \& $\gamma_{N_{c}}$}
\label{sec:exp}
\begin{figure*} [t]
\centering
\includegraphics[width=0.48\textwidth]{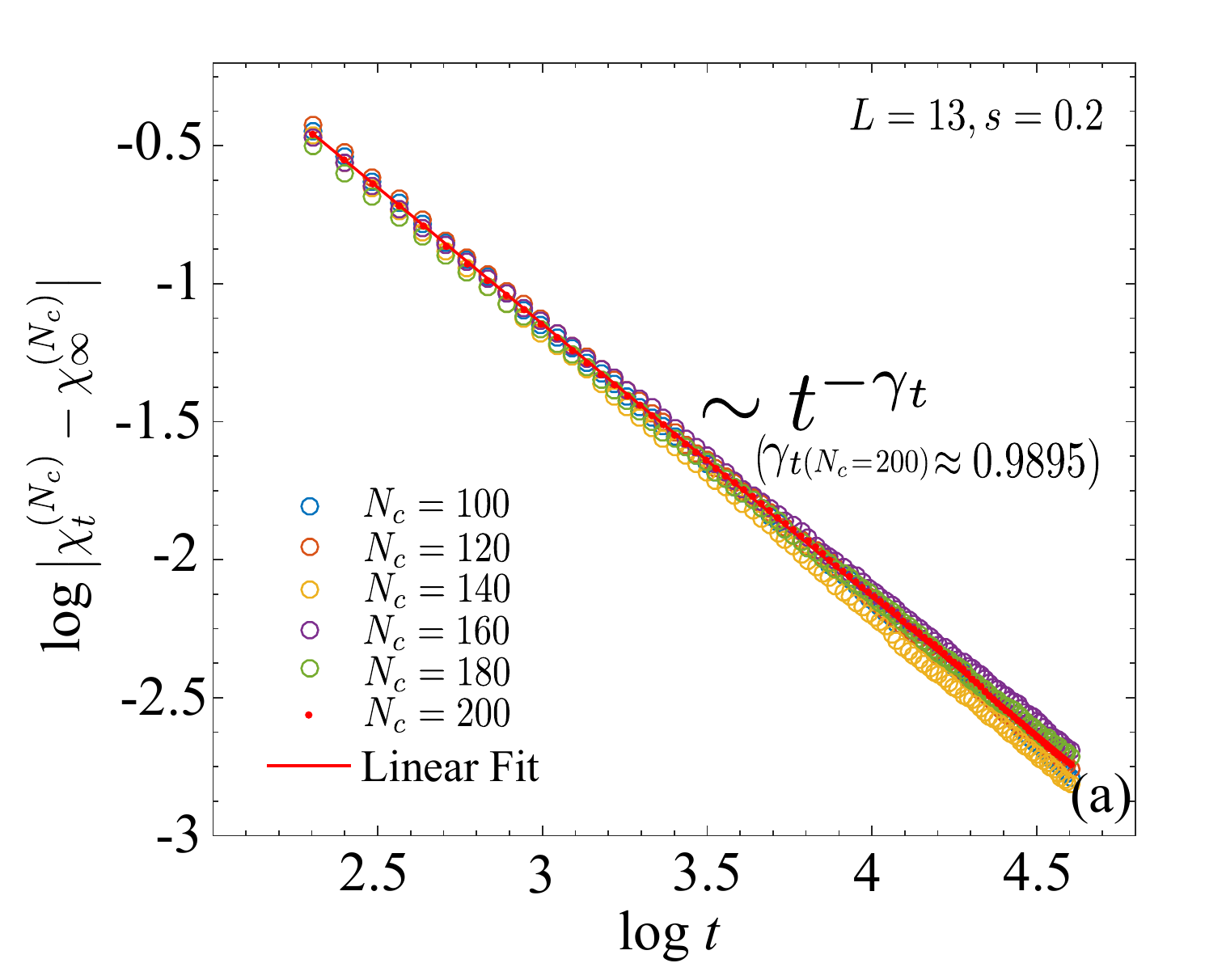}
\includegraphics[width=0.48\textwidth]{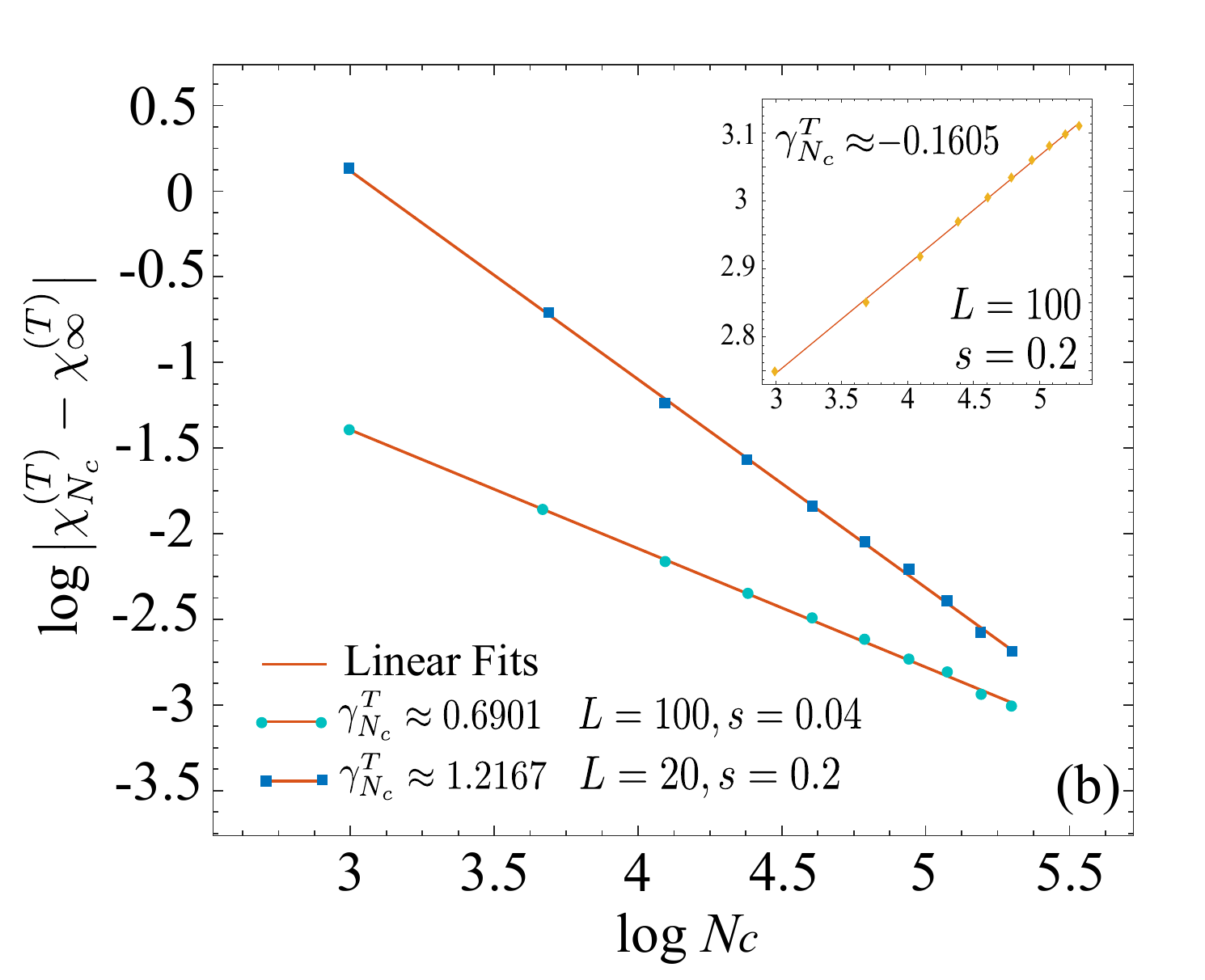}
\centering
\caption{\label{fig:powerlaw} Power law behavior of the generalized \textbf{(a)} finite-$t$ and \textbf{(b)} finite-$N_c$  scalings of the CGF estimator for a contact process with $\lambda=1.75$ and $h = 0.1$. The exponents  $\gamma_{N_{c}}$ and $\gamma_{t}$ 
were determined from the slope of a linear fit in log-log scale over Eqs.~(\ref{eq:nScal3}) and ~(\ref{eq:tScal2PL}), respectively. For $\gamma_{t}$, we used $L=13$, $s=0.2$ and $N_c \in \vec{N_c} = \{100,120,...,200 \}$. Meanwhile,  $\gamma_{N_{c}}$ was computed for $L=20$ and $s=0.2$, and for $L=100$ and $s=0.04$, which are contrasted with the one for $L=100$ and $s=0.2$ (inset) illustrating the change in the $N_c$-scaling depending on $s$ and $L$. Additionally, in all the cases $T=100$ and $R = 500$. 
}
\end{figure*}

From Eqs.~(\ref{eq:tScal2}) and~(\ref{eq:nScal2}) we expect that, independently of $N_c$, $T$, $L$ or $s$, a power law behavior of the form
\begin{eqnarray}
\label{eq:tScal2PL}
\big \vert \chi_{t}^{(N_{c})} - \chi_{\infty}^{(N_{c})} \big \vert &\sim  t^{-\gamma_{t}}, \\
\label{eq:nScal2PL}
\big \vert \chi_{\infty}^{(N_c)} - \chi_{\infty}^{\infty} \big \vert &\sim N_c^{-\gamma_{N_{c}}},
\end{eqnarray}
be observed. Thus, the exponents $\gamma_{t}$ \& $\gamma_{N_{c}}$ can be obtained from the slope of a straight curve in log-log scale of Eqs.~(\ref{eq:tScal2PL}) and~(\ref{eq:nScal2PL}) as can be seen in Fig.~\ref{fig:powerlaw}. Despite only some representative configurations have been presented, we confirm this power law behavior independently of the parameters chosen. 
Apart from characterizing the finite-$t$ and finite-$N_c$ behavior of the CGF estimator, the exponents $\gamma_{t}$ \& $\gamma_{N_{c}}$ provide valuable information about its convergence (or not) in the infinite-$t$ and infinite-$N_c$ limits. This convergence depends on the positivity of these exponents. However, in some cases they can take negative values (inset Fig.~\ref{fig:powerlaw}(b)) implying that the estimation lacks of an asymptotic limit becoming no longer reliable.
%

\section{Finite Scalings for a Large-$L$ Contact Process}
\label{sec: CPL100}
\begin{figure*} [t!]
\includegraphics[width=0.48\textwidth]
{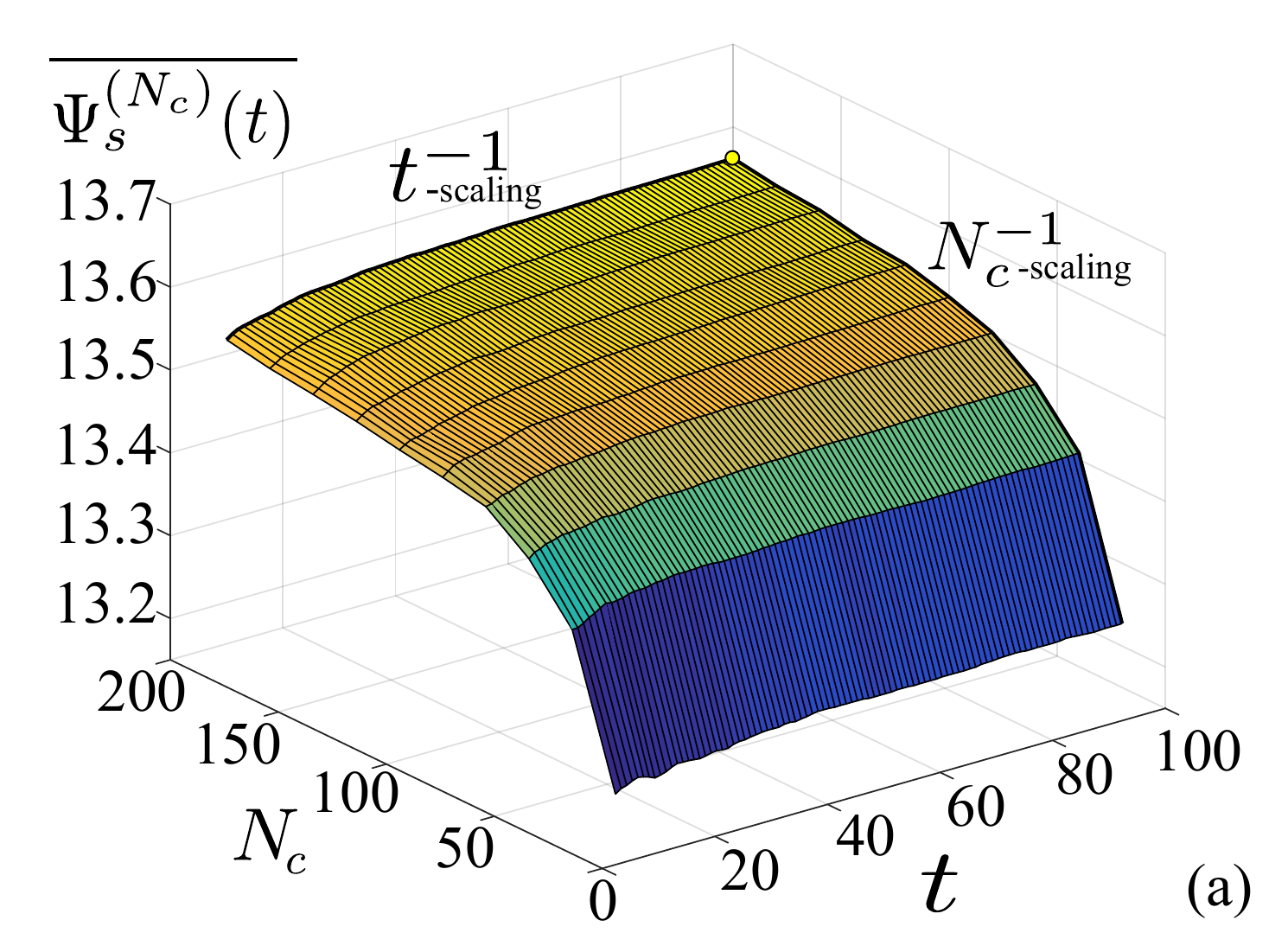}
\includegraphics[width=0.48\textwidth]
{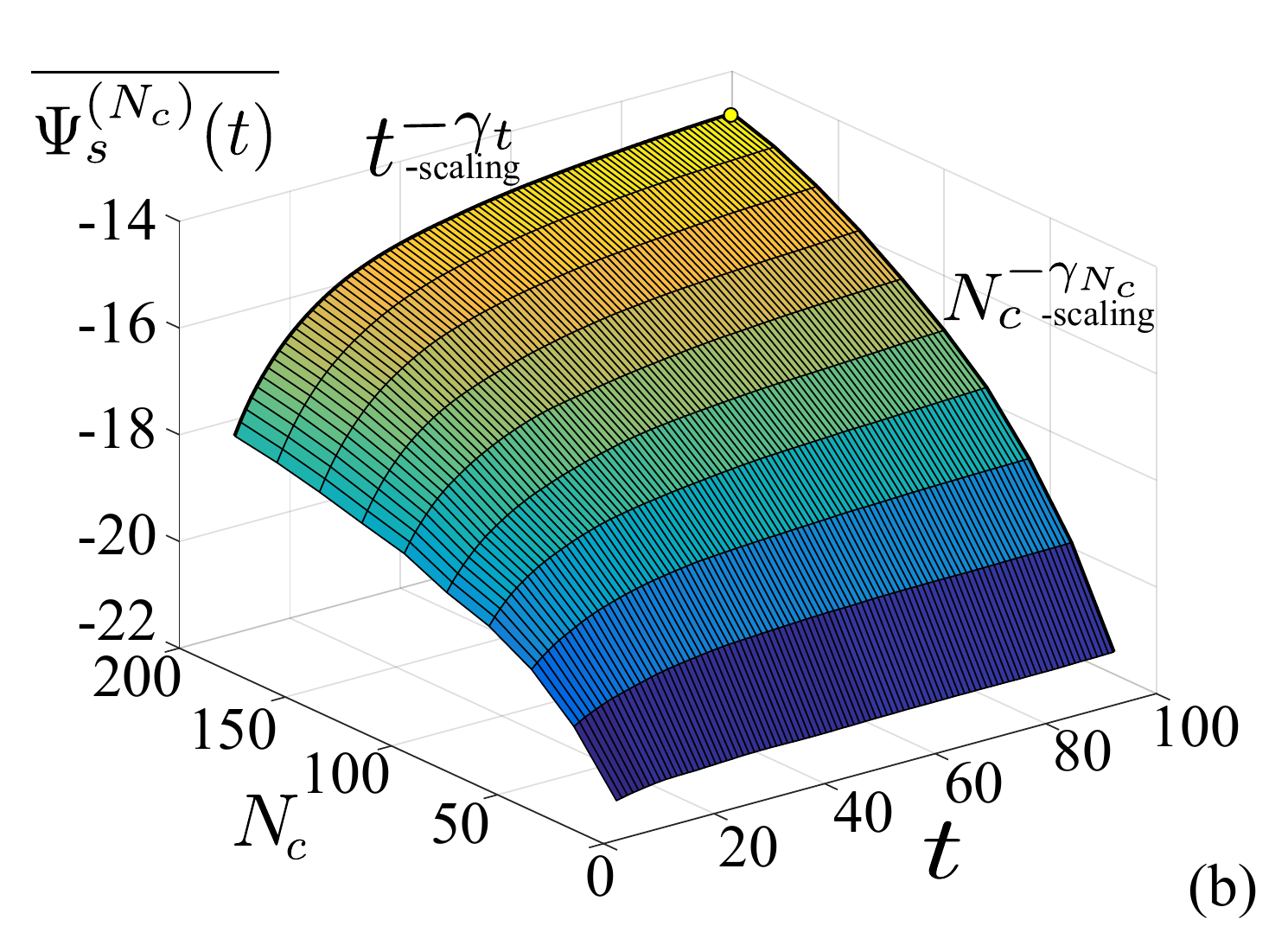}
\caption{\label{fig:surfPSI} CGF estimator  $\overline{ \Psi_{s}^{(N_{c})} (t) }$ (Eq.~(\ref{eq:PSI1})) as a function of time $t$ and the number of clones $N_c$ for \textbf{(a)} $s = -0.1$ and \textbf{(b)} $s = 0.2$. These surfaces were computed using the continuous-time cloning algorithm up a final simulation time $T=100$, $\vec{N}_{c} = \{20,...,200 \}$ and $R =500$ realizations for a contact process with $L=100$, $\lambda=1.75$ and $h = 0.1$. The $N_{c}^{-1}$-scaling observed in small-$L$ systems holds only for $s =-0.1$ whereas for $s=0.2$ a $N_{c}^{-\gamma_{N_{c}}}$-scaling is observed ($\gamma_{N_{c}}(s=0.2) \approx -0.16$). Similarly, for the time-scaling for which $\gamma_{t}(s=0.2) \approx 0.7$.
}
\end{figure*}

In Fig.~\ref{fig:surfPSI}, we compare the behavior of $\overline{\Psi_{s}^{(N_{c})}(t)}$ as function of $t$ and $N_{c}$, for two representative values of the parameter $s$, $s=-0.1$ (left) and $s=0.2$ (right). The size of the system is $L=100$ sites. Each point of these surfaces was obtained using the cloning algorithm (Eq.~(\ref{eq:PSI2})) up to time $T=100$, for $\vec{N_c} = \{20,40,...,180,200 \}$ and for $R=500$ realizations. 
The best possible CGF estimation 
(i.e., at largest $T$ and $N_c$)
in both cases is shown with solid circles
which, according to 
Ref.~\cite{partII}, could be improved by using the $t^{-1}$ and $N_c^{-1}$-scalings (if still valid for large-$L$).
\subsection{Finite-Time and Finite-$N_c$ Scalings}
\label{sec: CGFL100}

Although the exponents $\gamma_{t}$ and $\gamma_{N_{c}}$ can be  computed in principle for any value of $N_c \in \vec{N_c}$ and for any $t \leq T$, as we saw above, 
%
from now on, we will consider these exponents defined at the highest number of clones and at final simulation time, i.e.,
\begin{eqnarray}
\label{eq:GammaT}
\gamma_{t} &:= \gamma_{t}(N_{c}=\max \vec{N_{c}}), \\
\label{eq:GammaN}
\gamma_{N_c} &:= \gamma_{N_c}(t = T).
\end{eqnarray}
Thus, the exponent $\gamma_{t}$ is obtained as described in Sec.~\ref{sec:exp} after adjusting Eq.~(\ref{eq:tScal2}) to $\overline{ \Psi_{s}^{(N_c)}(t)}$ for $N_c=\max \vec{N_{c}}=200$. 
On the other hand, $\gamma_{N_{c}}$ is determined after fitting $\chi_{\infty}^{(N_c)}$ with Eq.(\ref{eq:nScal2}) at $T=100$ or, as $\gamma_{N_{c}} \approx \gamma_{N_{c}}^{T}$, after fitting $\overline{ \Psi_{s}^{(N_c)}(t=T)}$ using Eq.~(\ref{eq:nScal3}).
In simple words, these exponents can be obtained from an adequate fit over the thick curves in Fig.~(\ref{fig:surfPSI}). They characterize the finite-$t$ and finite-$N_c$ behavior of the large deviations of the dynamical activity $K$. 

Following this approach, we found that the $t^{-1}$-scaling~(\ref{eq:tScal1}) is satisfied only for $s=-0.1$, 
%
meaning that the exponent $\gamma_{t}$ was found to be $\gamma_{t} \approx 1$. As a consequence, the parameter $\chi_{\infty}^{(N_{c})}$ obtained from Eq.~(\ref{eq:tScal2}) effectively represents the limit in $t \to \infty$ of the CGF estimator, i.e., $\chi_{\infty}^{(N_{c})} \approx f_{\infty}^{(N_{c})}$. This is not the case for $s=0.2$ for which $\gamma_{t}(s=0.2) \approx 0.7$.
Similarly, 
a $N_{c}^{-\gamma_{N_{c}}}$-scaling is observed for $s = 0.2$, whereas for $s=-0.1$, the $N_c^{-1}$-scaling~(\ref{eq:nScal1}) holds. 
%
It is important to remark that a positive value of exponent $\gamma_{N_{c}}$ still guarantees the convergence of the CGF estimator in the infinite-$N_{c}$ limit. However, even though $\gamma_{N_{c}}(t=10)>0$ at initial times, at final time $T$, the exponent is negative ($\gamma_{N_{c}}(t=T) \approx -0.16$), implying that the CGF estimator lacks of a limiting value in the infinite-$t$ infinite-$N_c$ limit.
%
%
%
Below, we present how the change in the scalings is produced for values of $s$ within the interval $s \in [-0.1,0.2]$.

\subsection{Exponents Characterization \& $s$-Dependence}
\label{sec: gamma_tn}
%
%
%
%
For $s<0$, the exponent $\gamma_{t}$ varies around $1$. 
However for $s>0$, 
$\gamma_{t}$ deviates slightly from $1$ decreasing with $s$ up to $\gamma_{t} \approx 0.7$ at $s=0.2$. 
In order to describe the behavior of this exponent, results convenient to define 
$s'$ as the value of the parameter $s \in [s_a,s_b]$ such that $\gamma_{t}(s < s') \approx 1$, i.e., until which the $t^{-1}$-scaling holds. Thus,
%
%
\begin{equation}
\gamma_{t}(L=100):
\label{eq:Gammat_s}
\cases{\gamma_{t}(s) \approx 1    &for $s<s'$\\
            0<\gamma_{t}(s) <1    &otherwise.}
\end{equation}
If the scaling holds $\forall s \in [s_a,s_b]$ (given some system size $L$), then $s' = s_b$. 

\begin{figure*} [t]
\centering
\includegraphics[width=0.6\textwidth]{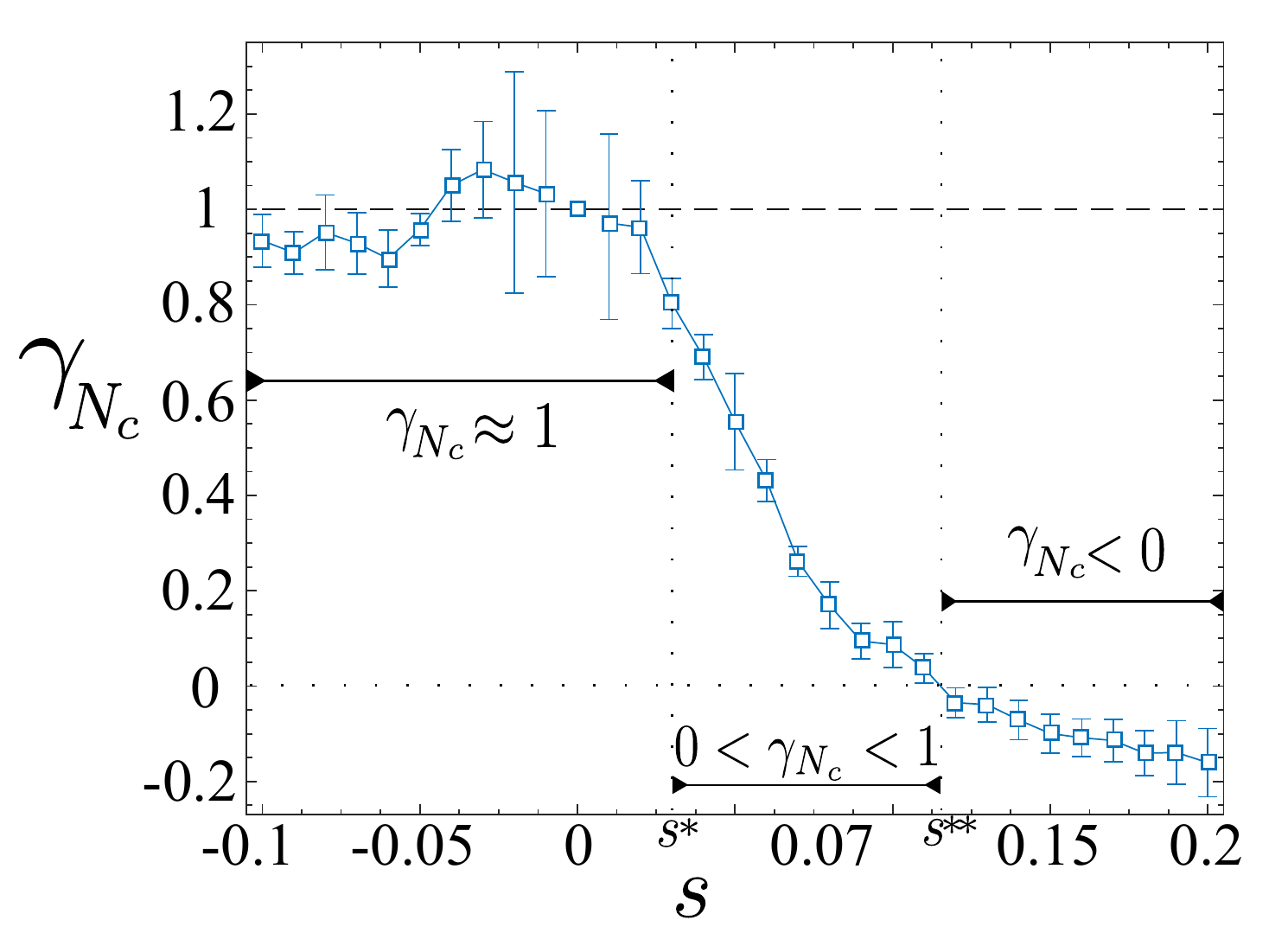}
\centering
\caption{\label{fig:Gamma_tn_s} Dependence of the $N_c^{-\gamma_{N_c}}$-scaling with the parameter $s \in [-0.1,0.2]$. 
The exponent $\gamma_{N_{c}}(s)$ is obtained by fitting the CGF estimator $\overline{\Psi_{s}^{(N_{c})}(T=100)}$ as function of $N_c \in \vec{N_c} = \{20,40,...,180,200 \}$ by Eq.~(\ref{eq:nScal3}) in log-log scale as described in Sec.~\ref{sec:exp}.
Three stages can be clearly distinguish for $\gamma_{N_{c}}(s)$ with $L=100$: i) $\gamma_{N_{c}}(s<s^{*})\approx 1$, ii) $0 < \gamma_{N_{c}}(s^{*} < s < s^{**})<1$, and iii) $\gamma_{N_{c}}(s>s^{**}) < 0$. The exponent $\gamma_{N_{c}}$ for $s=0$ was set to $\gamma_{N_{c}}(s=0) = 1$.
The error bars correspond to the $95\%$ confidence bounds on the coefficient $\gamma_{N_{c}}(s)$ associated to the fit. The goodness of fit with $R^{2}\gtrsim 0.999$ is valid for most values of $s$, except for the ones closer to $0$ for which $R^{2} \gtrsim 0.994$.
}
\end{figure*}

On the other hand, the value of $s \in [s_a,s_b]$ which signals the validity of the $N_{c}^{-1}$-scaling is denoted by $s^{*}$.
From this point, $\gamma_{N_{c}}$ decreases until eventually it becomes negative, as can be seen in Fig.~\ref{fig:Gamma_tn_s}. Here, we introduce $s^{**}$ such that $\gamma_{N_{c}}(s = s^{**}) = 0$ and thus $\gamma_{N_{c}}<0$ for $s > s^{**}$. 
This behavior was not observed for $L=6$ for which the $N_{c}^{-1}$-scaling was valid for all $s$~\cite{partII} and in those cases,  $s^{*}=s_b$ and $\nexists s^{**}$.
Here instead, we have distinguished clearly three stages for the exponent $\gamma_{N_{c}}(s)$:
%
%
\begin{equation}
\gamma_{N_{c}}(L=100):
\label{eq:GammaN_s}
\cases{\gamma_{N_{c}}(s)\approx 1    &for $s<s^{*}$\\
0<\gamma_{N_{c}}(s)<1,               &for $s^{*}<s<s^{**}$\\
            \gamma_{N_{c}}(s) < 0,   &for $s>s^{**}$.}
\end{equation}

The convergence of the CGF estimator to an asymptotic value ($f_{\infty}^{\infty}$) in the infinite-$t$ and infinite-$N_c$ limit relied not (necessarily) on the validity of the $t^{-1}$- and $N_{c}^{-1}$-scalings but on the positivity of the exponents $\gamma_{t}$ and $\gamma_{N_{c}}$. However, as presented in this section (and as it can be seen in Fig.~\ref{fig:Gamma_tn_s}), the exponent $\gamma_{N_{c}}(L=100)$ takes negative values for $s>s^{**}$ implying a lack of a limiting value in this region. Below we present how these asymptotic values are affected as the exponents $\gamma_{t}$ and $\gamma_{N_{c}}$ change with $s$.

\newpage

\subsection{Scaling Method and Asymptotic CGF Limits}
\label{sec: SML100}
\begin{figure} [t]
 \centering
\includegraphics[width=0.6\textwidth]{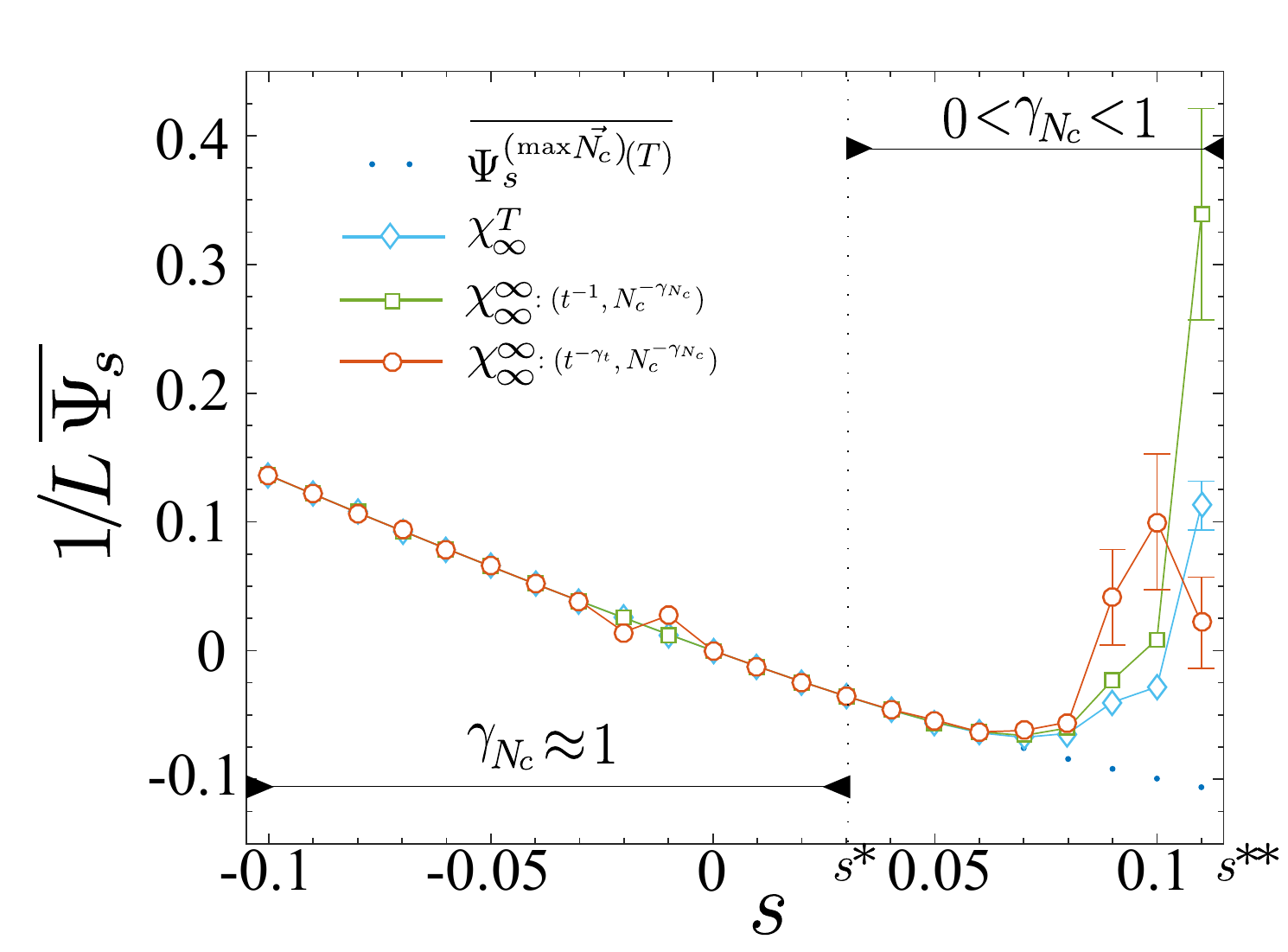}
 \centering
\caption{\label{fig:Psi_s} Different estimators of the large deviations of the activity as function of the parameter $s \in [-0.1,0.1]$ for a contact process with $L=100$ sites. The standard CGF estimator $\overline{\Psi_{s}^{(N_{c})}(t)}$ (evaluated at $N_c = \max \vec{N_c}=200$, $t = T=100$ and for $R=500$ realizations) is shown with dots meanwhile the asymptotic limits obtained from the scaling method $\chi_{\infty}^{\infty}$ are presented in squares and circles, and $\chi_{\infty}^{T}$ in diamonds. The legend $(t^{-1},N_{c}^{-\gamma_{N_c}})$ refers to the assumption of a $t^{-1}$-scaling for $\overline{\Psi_{s}^{(N_{c})}(t)}$ (setting $\gamma_{t}=1$ in Eq.~(\ref{eq:tScal2})) and a $N_{c}^{-\gamma_{N_c}}$-scaling~(\ref{eq:nScal2}) for the $\chi_{\infty}^{(N_c)}$'s. On the other hand, $(t^{-\gamma_{t}},N_{c}^{-\gamma_{N_c}})$ refers to the fact that we have left $\gamma_{t}$ and $\gamma_{N_c}$ as free parameters.
The different estimators correspond to each others up to $s=s^{*}$ from which they diverge up to $s = s^{**}$. This is directly related with the behavior of the exponent $\gamma_{N_c}$ observed in Fig.~\ref{fig:Gamma_tn_s}(b). The error bars correspond to the $95\%$ confidence bounds on the coefficients $\chi(s)$ associated to the fit. The goodness of fit with $R^{2}\gtrsim  0.995$ is valid for most of the values of $s$ except for ones closer to $s^{**}$. Only for these values, the errors bars are presented (re-scaled to the $10\%$), for the rest of values the re-scaled error bars are of the order of the size of the marker or smaller.
}
\end{figure}

The scaling method~(see \ref{sec:AppSM}) allows to determine the asymptotic limit to which the CGF estimator~(\ref{eq:PSI1}) converges in the $t \to \infty$ and $N_c\to \infty$ limits. Moreover, this limit, that we have denoted $f_{\infty}^{\infty}$ (Eq.~(\ref{eq:nScal1})), was proved to render a better estimation of the analytical CGF $\psi(s)$ than the standard estimator $\overline{ \Psi_{s}^{(\max \vec{N_c})}(T)}$, at least for the cases analyzed in Ref.~\cite{partII}.
However, the evidence we just presented 
would suggest that the determination and existence of $f_{\infty}^{\infty}$ depend
on the values of the exponents $\gamma_{t}$ and $\gamma_{N_{c}}$. 
Thus, an asymptotic limit for the CGF estimator 
exists only for $s<s^{**}$ (for which the exponents $\gamma_{t}$ and $\gamma_{N_{c}}$ are positive) and only for $s<s^{*}$ (for which $\gamma_{t} \approx \gamma_{N_{c}} \approx 1$) 
the extracted $\chi_{\infty}^{\infty}$ (obtained from Eq.~(\ref{eq:nScal2}))
corresponds to $f_{\infty}^{\infty}$. 
Indeed, this can be observed in Fig.~\ref{fig:Psi_s} where we have applied the scaling method to our example. 

The method can be performed following two different approaches: 
\textit{i)} $(t^{-1},N_{c}^{-\gamma_{N_c}})$: First, imposing a $t^{-1}$-scaling for $\overline{\Psi_{s}^{(N_{c})}(t)}$ (setting $\gamma_{t}=1$ in Eq.~(\ref{eq:tScal2})) and then, considering a $N_{c}^{-\gamma_{N_c}}$-scaling~(\ref{eq:nScal2}) for the extracted $\chi_{\infty}^{(N_c)}$'s. Alternatively,
\textit{ii)} $(t^{-\gamma_{t}},N_{c}^{-\gamma_{N_c}})$: Leaving $\gamma_{t}$ and $\gamma_{N_c}$ as free parameters in Eqs.~(\ref{eq:tScal2}) and~(\ref{eq:nScal2}). 
Both resulting estimators $\chi_{\infty}^{\infty}(i)$ and $\chi_{\infty}^{\infty}(ii)$ are shown in Fig.~\ref{fig:Psi_s} with squares and circles, respectively. Additionally, the infinite-$N_c$ limit $\chi_{\infty}^{T}$~(\ref{eq:nScal3}) is also presented with diamonds. 
The standard CGF estimator $\overline{\Psi_{s}^{(\max \vec{N_c})}(T)}$ (in dots) serves as reference. 

\newpage
As can be seen in Fig.~\ref{fig:Psi_s}, the different estimators correspond to each others up to $s=s^{*}$.
%
%
From this point, their value and distance with respect to $\overline{\Psi_{s}^{(\max \vec{N_c})}(T)}$ increase rapidly with $s$ up to $s = s^{**}$.
%
%
This behavior 
keeps correspondence with the $N_c^{-\gamma_{N_c}}$-scaling of the CGF estimator. Specifically, with the stages of the exponent $\gamma_{N_c}$ that were presented in Sec.~\ref{sec: gamma_tn} and Fig.~\ref{fig:Gamma_tn_s}. 
Thus, the lack of an asymptotic limit is related precisely with the change in sign of $\gamma_{N_c}$ in $s = s^{**}$ in the same way as the divergence of the estimators from the standard one at $s=s^{*}$ is related with the fact that from this point, $\gamma_{N_c} \neq  1$.

The example presented through this section related the existence of an asymptotic limit to which the CGF estimator converges with its actual finite scalings in large-$L$ systems. 
Below we extend our analysis by considering the scaling behavior on a wider range of values of $L$. This will provide a complete overview of how the CGF estimator behaves and how the change in scaling is given.

\section{$L$-Dependence of the Finite Scalings}
\label{sec: planeSL}

In this section, we detail the behavior of the finite-$t$ and -$N_c$ scalings of the CGF estimator 
for $s>0$ and $L$ ranging in the interval $L\in [3,100]$. For each pair $(s,L)$, the exponents $\gamma_{t}$ and $\gamma_{N_c}$ were computed as described in Sec.~\ref{sec:exp} for $T=100$ and $\vec{N_c} = \{20,40,...,180,200 \}$. 

\begin{figure} [t]
\centering
\includegraphics[width=0.6\textwidth]{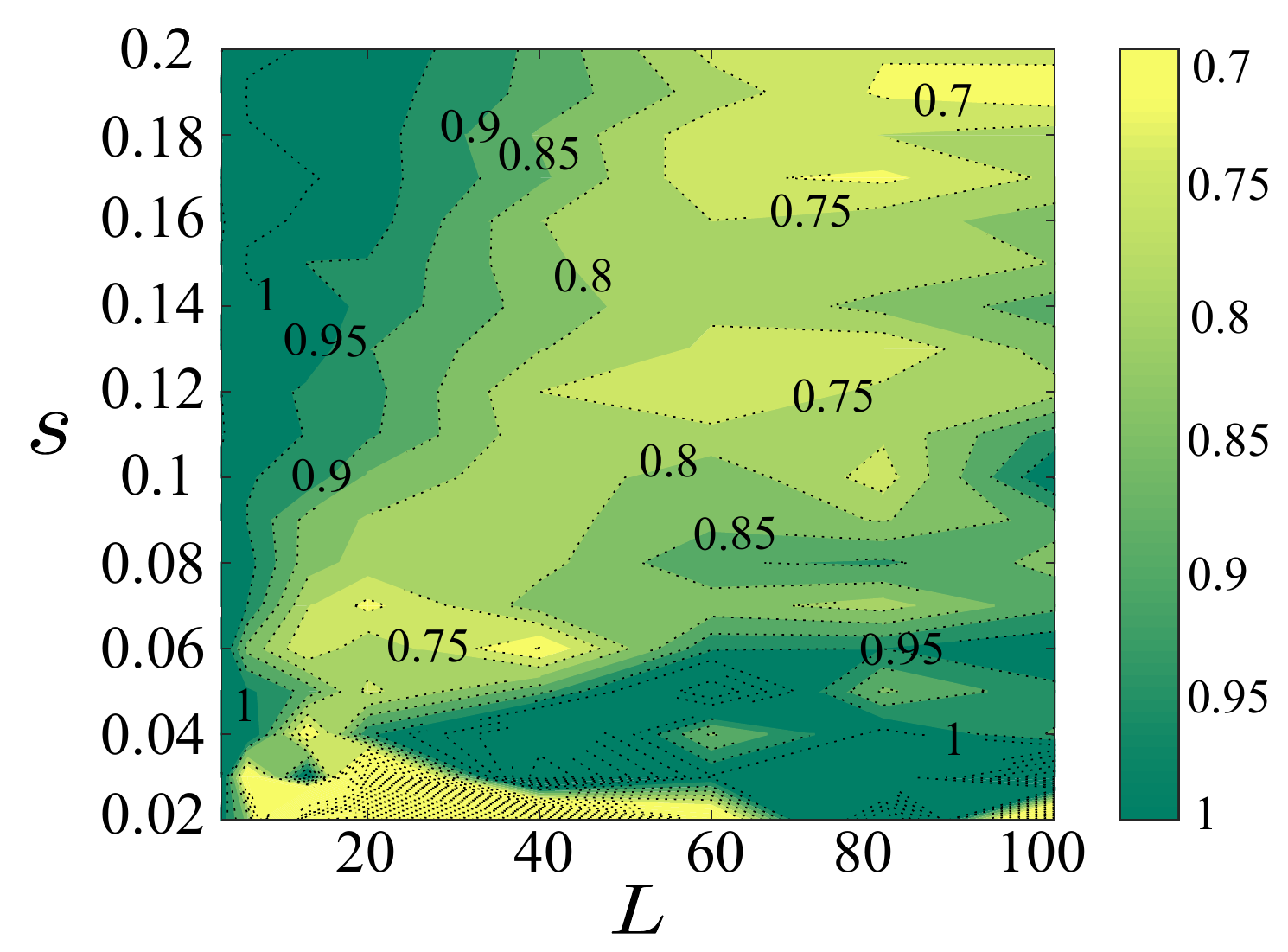}
\centering
\caption{\label{fig:Gammat_LS} Behavior of the exponent $\gamma_{t}$ in a region of the plane $s-L$ given by the parameters $s \in \big[0.02,0.2\big]$ and $L\in \big[ 3,100 \big]$. This exponent characterizes the finite-time scaling of the large deviations of the activity in the contact process (\textbf{$t^{-\gamma_{t}}$-scaling}). 
The values of $\gamma_{t}$ closest to $1$ are presented with the darkest tones whereas smaller values are shown with clearer ones.
Although for small values of $L$ the $t^{-1}$-scaling holds independently of $s$, in general the exponent $\gamma_{t}$ decreases gradually as $s$ and $L$ increase.
}
\end{figure}

\subsection{Characterization of the exponent $\gamma_{t}(s,L)$}
The contour plot in Fig.~\ref{fig:Gammat_LS} shows the value of the exponent $\gamma_{t}$ as it changes depending on the parameters $s$ and $L$. 
%
%
We have focused in the region for $s\in \big[0.02,0.2\big]$ as for $s<0$, $\gamma_{t}\approx 1$ and thus, the $t^{-1}$-scaling~(\ref{eq:tScal1}) holds. 
The values closest to $1$ are presented with the darkest tone while smaller values are shown with clearer tones. As can be seen, the exponent $\gamma_{t}$ decreases gradually as $L$ and $s$ increase.

For a given system size $L$, we can describe qualitatively the behavior of $\gamma_{t}$ with respect to $s$ is similar way as we did for $L=100$ in Sec.~\ref{sec: gamma_tn}.  
%
In order to extend that description into the plane $s-L$, we introduce a number of sites dependency of the bound $s'$. We denote by $s'(L)$ the value of $s$ until which the $t^{-1}$-scaling is valid given a particular $L$. Similarly, $\gamma_{t}^{\circ}(L)$ is the lower bound of $\gamma_{t}^{(L)}(s)$.
%
%
%
Thus, the exponent $\gamma_{t}$ which characterizes the $t^{-\gamma_{t}}$-scaling~(\ref{eq:tScal2}) of the CGF estimator 
%
%
is given by
%
%
%
%
%
\begin{equation}
\gamma_{t}:
\label{eq:Gammat_sL}
\cases{\gamma_{t}^{(L)}(s) \approx 1,                              &for $s<s'(L)$\\
\gamma_{t}^{\circ}(L) \leq \gamma_{t}^{(L)}(s) \lesssim 1,     &otherwise }
\end{equation}
where $s'(L)>0$, $\gamma_{t}^{\circ}(L)>0$ and $L$ is large. In fact, for this case, $\gamma_{t}^{\circ}(L)>1/2$, for all $L$.

\subsection{Characterization of the exponent $\gamma_{N_{c}}(s,L)$}

\begin{figure*} [t]
\centering
\includegraphics[width=0.48\textwidth]{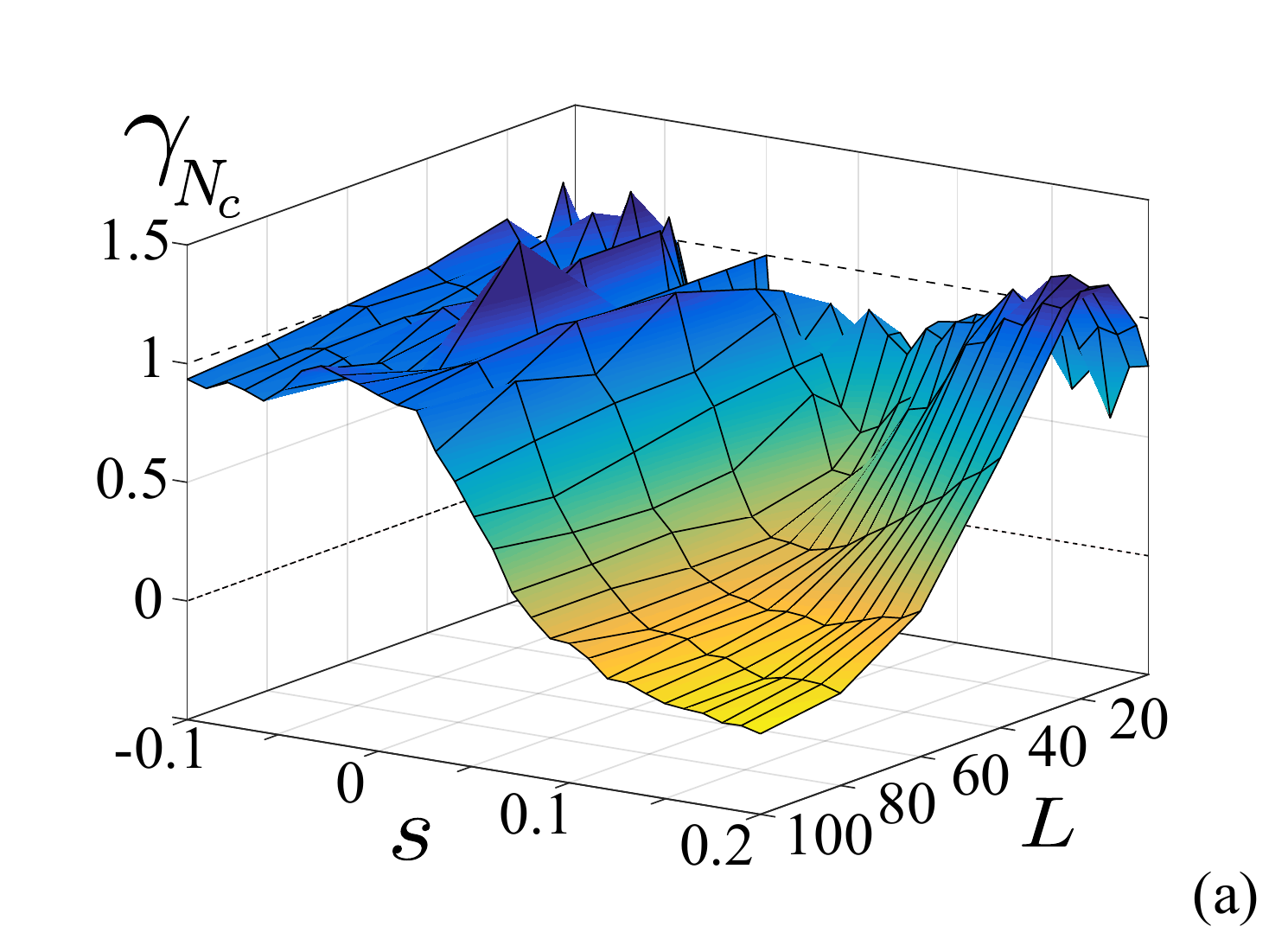}
\includegraphics[width=0.48\textwidth]{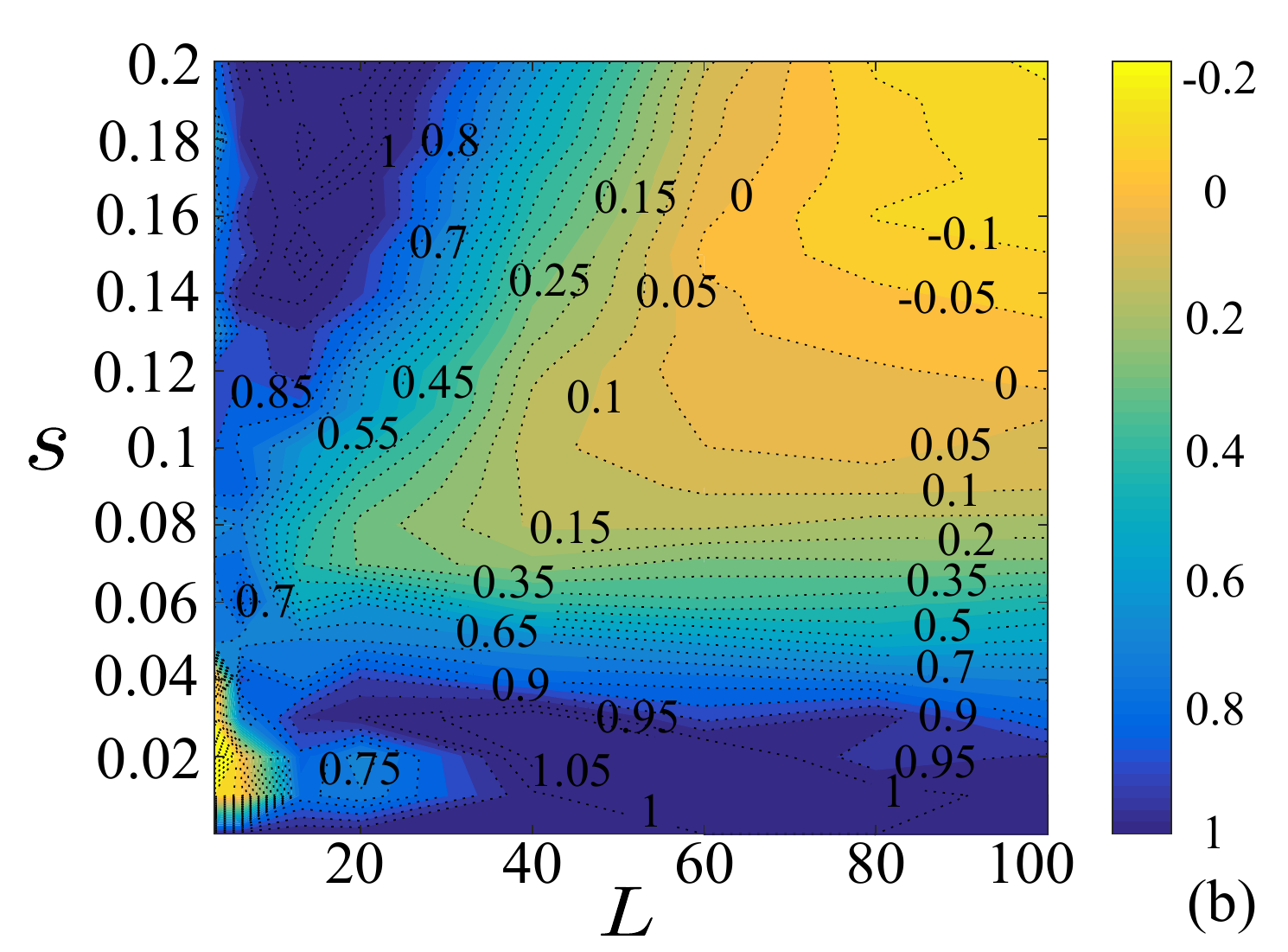}
\caption{\label{fig:Gamman_LS} Behavior of the exponent $\gamma_{N_c}$ in a region of the plane $s-L$ given by the parameters $s \in \big[-0.1,0.2\big]$ and $L\in \big[ 3,100 \big]$. This exponent characterizes the finite-$N_c$ scaling of the large deviations of the activity in the contact process (\textbf{$N_c^{-\gamma_{N_c}}$-scaling}). \textbf{(a)} The surface $\gamma_{N_c}(s,L)$ illustrates the change in the scaling of the CGF estimator. The exponent $\gamma_{N_{c}}(s,L) \approx 1$  up to some value of $s$ (which depends of $L$), from which it decreases as $s$ and $L$ increases, even becoming negative for some values of $L$.  \textbf{(b)} Projection of the surface in (a) on the plane $s-L$. The values of $\gamma_{N_{c}}(s,L)$ closer to $1$ are shown in dark tones while the smaller tones with clearer ones. The $N_c^{-\gamma_{N_c}}$-scaling can be characterized depending on the number of zeros of the exponent $\gamma_{N_{c}}(s)$ for a given $L$ in three regions: $\mathcal{L}_{I}$ if has no zeros, $\mathcal{L}_{II}$: if have two zeros and $\mathcal{L}_{III}$: if have one zero. 
}
\centering
\end{figure*}

Similarly as above, in Fig.~\ref{fig:Gamman_LS} we present the exponent $\gamma_{N_{c}}$ as it changes depending of some particular choice of the parameters $(s,L)$ within the intervals considered. 
The surface in Fig.~\ref{fig:Gamman_LS}(a) illustrates clearly the change in the $N_c$-scaling of the CGF estimator.
For every value of $L$ considered, the exponent $\gamma_{N_{c}}$ is approximately $1$ up to some value of $s$, denoted as $s^{*}(L)$ (Sec.~\ref{sec: gamma_tn}). 
%
However, from this point, its value decreases as $s$ and $L$ increases, becoming, in some cases, negative.
This change in the $N_c^{-\gamma_{N_c}}$-scaling is also shown in the contour plot in Fig.~\ref{fig:Gamman_LS}(b) where we have focus in the region for $s>0$. The values of $\gamma_{N_{c}}$ closer to $1$ are shown in dark tones. 

In Sec.~\ref{sec: gamma_tn}, we also defined $s^{**}$ such that $\gamma_{N_{c}}^{(L)}(s^{**}) = 0$. This value of course depends on $L$ and in some cases it does not even exists.
However, for some particular values of $L$ (large), the exponent $\gamma_{N_{c}}^{(L)}$ changes sign twice (as can be seen in Fig.~\ref{fig:Gamman_LS}(b)).
We will use this fact in order to characterize the $N_c^{-\gamma_{N_c}}$-scaling depending on the number of zeros of the exponent $\gamma_{N_{c}}^{(L)}(s)$ for a given $L$.
We define $\mathcal{L}_{I}$ as the set of values of $L$, for which the exponent $\gamma_{N_{c}}^{(L)}(s)$ has no zeros, $\mathcal{L}_{II}$: if has two zeros ($s_{1}^{**}(L)$ and $s_{2}^{**}(L)$, with $s_{2}^{**}(L) > s_{1}^{**}(L)$) and $\mathcal{L}_{III}$: if has one zero  ($s^{**}(L)$). 
These regions are bounded by $L_{inf}$ and/or by $L_{sup}$, where $L_{inf}$ is the smallest value of $L$ such that the curve $L =L_{inf}$ is tangent to $\gamma_{N_{c}}(s,L)=0$ in one single point. On the other hand, $L_{sup}$ is the largest $L$ such that the curve $L = L_{sup}$ cuts $\gamma_{N_{c}}(s,L)=0$ in two points. 
Thus, the region $\mathcal{L}_{I}$ groups the values of $L$ such that  $L<L_{inf}$, $\mathcal{L}_{II}$ the values of $L$ within the interval $L_{inf} < L < L_{sup}$ and $\mathcal{L}_{III}$, the values of $L$ such that $L > L_{sup}$.
%
Thus, the exponent $\gamma_{N_{c}}$ which characterizes the $N_c^{-\gamma_{N_c}}$-scaling~(\ref{eq:nScal2}) of the CGF estimator is given by

\begin{equation}
\label{eq:GammaN_sL}
\gamma_{N_{c}}:
\cases{
\mathcal{L}_{I}: \cases{
    \gamma_{N_{c}}^{(L)}(s)\approx 1, &for $s<s^{*}(L)$\\
    0 < \gamma_{N_{c}}^{(L)}(s) \lesssim 1, &otherwise.   
          } \\
\mathcal{L}_{II}: \cases{
    \gamma_{N_{c}}^{(L)}(s)\approx 1, &for $s<s^{*}(L)$\\
     0 < \gamma_{N_{c}}^{(L)}(s) < 1, &for $s^{*}(L)<s<s^{**}_{1}(L)$ \\
                                      &and $s>s^{**}_{2}(L)$\\                                                                  
    \gamma_{N_{c}}^{(L)}(s)<0,  &for $s^{**}_{1}(L)<s<s^{**}_{2}(L)$
          } \\
\mathcal{L}_{III}: \cases{
    \gamma_{N_{c}}^{(L)}(s)\approx 1, &for $s<s^{*}(L)$\\
    0 < \gamma_{N_{c}}^{(L)}(s) < 1, &for $s^{*}(L)<s<s^{**}(L)$\\
    \gamma_{N_{c}}^{(L)}(s)<0,  &for $s>s^{**}(L)$  
          } \\
}
\end{equation}


\section{Dynamical Phase Transition, Scalings and the Contact Process}
\label{sec:DPTcp}

In Sec.~\ref{sec:LD}, we introduced the biasing parameter (or field) $s$ (conjugated to an observable $\mathcal O$) in order to characterize a non equilibrium ensemble of trajectories. 
%
%
Within this ``$s$-ensemble'', space-time or \textbf{dynamical phase transitions} 
manifest themselves as singularities in the CGF and, in our case, express a dynamical coexistence of histories with high and low activity $K$~\cite{Lecomte2007}.

The contact process~\cite{CP, GRASSBERGER1979373, liggett2012interacting} is well know to exhibit a dynamical phase transition in the $L \rightarrow \infty$ limit~\cite{lecomte_numerical_2007, Lecomte2007, ThermoCP, marro_dickman_1999} even in one-dimension~\cite{marro_dickman_1999}. 
However in Ref.~\cite{lecomte_numerical_2007} evidence of the presence of a phase transition (in the active phase of $\lambda$) was reported to occur at $s_c\approx0.057$ for finite-$L$. There, the authors used the same version of the contact process and the same approach we used throughout this paper (i.e., the cloning algorithm).
On the other hand, in Ref.~\cite{ThermoCP}, using a 
density matrix re-normalization group approach (DMRG)~\cite{WhiteDMRG1, WhiteDMRG2, reviewDMRG, Kaulke1998, Carlon1999}, 
it was showed that for every value of infection rate $\lambda$, either if this belong to the absorbing or to the active phase, there exists a phase transition as a function of $s$. 
For the case of the active phase, this transition was found to occur at $s_c = 0$. 
It is important to remark that even if the versions of the contact process used in Refs.~\cite{lecomte_numerical_2007} and~\cite{ThermoCP} are different, both present a dynamical phase transition.  Meanwhile in the later case the particles are created just at the boundaries, in Ref.~\cite{lecomte_numerical_2007} (and here) they are created at every site
and also, the spontaneous rate of creation $h$ is considered different from $0$ (in order to circumvent the absorbing state in finite size~\cite{Lecomte2007}).

\newpage

Despite our main interest 
is not the study of the dynamical phase transition in the contact process, 
what does concern us is how this could affect the finite scalings and convergence of the CGF. Importantly, the relation that $s^{*}$ (but also $s'$ or $s^{**}$) could have with $s_c$ where this transition occurs. 
In Sec.~\ref{sec: Scaling} we showed how the scaling behavior given by Eqs.~(\ref{eq:tScal2}) and~(\ref{eq:nScal2}) was robust independently of $T$, $N_c$, $s$ or $L$ (Fig.~\ref{fig:powerlaw}), but not the exponents $\gamma_{t}$ and $\gamma_{N_c}$ whose behavior change depending on $s$ and $L$ specially in $s\geq0$ as $L$ becomes larger.
We remark that even if the infinite-$L$ limit is not achievable numerically,
the effects induced by a
dynamical phase transition should become more evident as $L$ increases  (which could explain many of the behavior observed throughout this paper).
This was clearly illustrated for $L=100$ for which $\gamma_{N_c}$ has an abrupt change for $s\geq0$, where we know the dynamical phase transition occurs, even taking negative values and inducing a divergence of the infinite-$t$ and infinite-$N_c$ limit of the CGF estimator (Fig.~\ref{fig:Psi_s}).

We recall here that our purpose was to verify the validity of the scalings (and thus, the convergence of the CGF estimator) presented in Ref.~\cite{partI,partII} (for small size systems) in the large-$L$ limit. 
A main feature of that study was the possibility of
extracting the infinite-$N_c$ infinite-$t$ limit of the CGF estimator from finite and small number of clones and simulation time. 
An analysis of the dynamical phase transition, on the other hand,
would require
a large-$N_c$ and -$t$ configuration which under our approach
is a task difficult to fulfill.
%
This however does not represent any surprise 
given that is well know that 
the existing methods~\cite{giardina_direct_2006, Hedges1309, PitardDT, SpeckDPT, Speck}
perform poorly in the vicinity of a dynamical phase transition, 
or they are numerically expensive in order to obtain accurate estimations~\cite{SpeckDPT, LimmerIce, DPTpath} developing if not important finite-size effects~\cite{hurtado_current_2009}.
However, recently has been proposed a promising method~\cite{nemoto_population-dynamics_2016, PhysRevLett.118.115702} which combines the existing cloning algorithm~\cite{giardina_direct_2006, giardina_simulating_2011, tailleur_simulation_2009, lecomte_numerical_2007, Hedges1309, PitardDT, SpeckDPT, Speck, partI, partII} with a modification of
the dynamics~\cite{jack_large_2010, 1742-5468-2010-10-P10007, PhysRevLett.111.120601, PhysRevLett.112.090602} resulting in a significant improvement of its computational efficiency. The method was successfully applied to the study of the dynamical phase transition of 1D FA model~\cite{FAmodel} using a relatively small $N_c$ and $L$. 
The implementation of this method will 
provide 
in a next stage
a clear contrast between the results obtained following the two different approaches and a correct relation between $s_c$ and $s^{*}$. 

\section{Conclusion}
\label{sec:conclusion}
In this paper, we analyzed the finite scalings of the large deviations of the activity in the contact process. We used the continuous-time version of the cloning algorithm from which the CGF (a estimator of LDF) can be obtained from the exponential growth (or decay) rate of a set of $N_c$ copies of the system which evolves following a modified dynamics up to a time $t=T$. 
It is expected that in the infinite-$t$ and -$N_c$ limits, this method provides an accurate CGF estimation. However, in practice, the best estimation is obtained from large  but finite simulation time $T$ and number of clones $N_c$. 

\newpage
The dependence of this estimator (and of its accuracy) with these two parameters was studied in Refs.~\cite{partI, partII}. The finite-$t$ and finite-$N_c$ scalings of the systematic errors of the CGF were found to behave as $1/N_c$ and $1/t$ in the large-$N_c$ and large-$t$ asymptotics, respectively. By making use of these convergence-speeds, it was proposed a (scaling) method which allowed to extract the asymptotic behavior of the CGF estimator in the $t \to \infty$ and $N_c \to \infty$ limits. At least for the cases analyzed in Refs.~\cite{partI, partII}, this infinite-time and infinite-$N_c$ limit resulted to render a better CGF estimation in comparison with the standard estimator. 
However, the validity of these scalings and thus, the convergence of
the estimator was proved only for a simple one-site annihilation-creation dynamics and for a contact process with $L=6$ sites, leaving an analysis of the dependence 
with the number of sites $L$ pending.

In order to do so, in this paper we redefined these scalings in a more general way by introducing the exponents $\gamma_{t}$ and $\gamma_{N_{c}}$. By doing this, we assumed the behavior of the CGF estimator described by a
$t^{-\gamma_{t}}$-scaling (Eq.~(\ref{eq:tScal2})) and a $N_{c}^{-\gamma_{N_{c}}}$-scaling (Eq.~(\ref{eq:nScal2}) from which the exponents 
can be obtained from the slope of a straight curve in log-log scale of Eqs.~(\ref{eq:tScal2PL}) and~(\ref{eq:nScal2PL}). 
These exponents not only 
characterize the finite-$t$ and finite-$N_c$ behavior of the CGF estimator for any system size $L$, but also provide valuable
information about its convergence. While these exponents take positive values, there is going to exist an asymptotic limit to which the estimator converges in $t \to \infty$ and $N_c \to \infty$, however we presented evidence showing this is not always true.

The CGF scaling analysis was done at first in Sec.~\ref{sec: CGFL100} where we considered a contact process with $L=100$ sites and two representative values of the parameter $s$.
Although the $t^{-1}$-scaling and $N_{c}^{-1}$-scaling were proved to hold for $s = -0.1$, this was not the case for $s=0.2$. 
%
%
%
Specifically, 
in Sect.~\ref{sec: gamma_tn}, we showed that the $N_{c}^{-1}$-scaling was valid up to $s=s^{*}$, then $\gamma_{N_{c}}$ decreases to $0$ at $s=s^{**}$ and finally, it becomes negative for $s>s^{**}$ implying that for this last  region the CGF estimation lacks of a limiting value in the infinite-$t$ infinite-$N_c$ limit and the estimation provided by the cloning algorithm is no longer reliable.
Indeed, the different asymptotic limits of the CGF estimator corresponded to each others up to $s=s^{*}$ from which they diverge up to $s = s^{**}$ (Sec.~\ref{sec: SML100}).
This analysis was extended to the plane $s-L$ in Sec.~\ref{sec: planeSL} where the exponents $\gamma_{t}$ and $\gamma_{N_c}$ were computed for a grid of values of the parameters $(s,L)$.
Their characterization was done introducing a number-of-sites dependency of the bounds $s'$, $s^{*}$ and $s^{**}$ previously defined in Sec.~\ref{sec: CPL100} as well as the use of the number of zeros of the exponent $\gamma_{N_{c}}^{(L)}(s)$ in order to characterize the different groups of $L$.
Whether the results presented through this paper are restricted only to the contact process or not is left as a pending problem and a possible direction for future research.

\newpage
\section*{Acknowledgements}
Esteban Guevara thanks Khashayar Pakdaman and Vivien Lecomte for their support and discussions. Special thanks to the Secretar\'{i}a Nacional de Educaci\'{o}n Superior, Ciencia, Tecnolog\'{i}a e Innovaci\'{o}n del Ecuador, SENESCYT.

\appendix

\section{Models and Methods}
\label{sec:App}
\subsection[\qquad \qquad \quad Contact Process]{Contact Process}
\label{sec:CP}
The process of interest throughout this paper consists in a one-dimensional lattice with $L$ sites and periodic boundary conditions known as contact process~\cite{CP, GRASSBERGER1979373, liggett2012interacting}. Each site $i$ in this lattice is occupied by a spin which can be in two possible states, $n_{i}=0$ or $n_{i}=1$, and with transition rates
\begin{eqnarray}
\label{eq:W_CP}
W(n_{i}=1 \rightarrow n_{i}=0) &= 1, \nonumber \\
W(n_{i} = 0 \rightarrow n_{i} = 1) &= \lambda(n_{i-1}+n_{i+1}) + h,
\end{eqnarray}
where $\lambda$ and $h$ are positive constants. The spontaneous rate of creation $h$ is introduced in the model in order to circumvent the absorbing state in finite size~\cite{Lecomte2007}. The contact process has been used to model the spread of infection diseases~\cite{10.2307/2244329}. Within this context, the state $n_{i}=1$ is used to represents a sick individual, and $\lambda$ can be seen as a  infection rate. 
The contact process is a model of the directed percolation universality class and its scaling properties have been discussed extensively~\cite{marro_dickman_1999, HinrichsenNoneq, OdorUniversality} and it is well known that it exhibits a dynamical phase transition in the $L \rightarrow \infty$ limit~\cite{lecomte_numerical_2007, Lecomte2007, ThermoCP, marro_dickman_1999}.

\subsection[\qquad \qquad \quad Cloning Algorithm]{Continuous-Time Cloning Algorithm}
\label{sec:AppCA}

Consider $N_{c}$ clones or copies of the system initially in the same configuration at ${\mathbf t}=0$. Each copy will evolve (continuously in time). The times in which this evolution occurs are denoted by $\mathbf t = \{ t^{(i)} \}_{i=1,...,N_{c}}$ and the configurations of the copies by $c = \{c_{i}\}_{i=1,...,N_{c}}$.
\begin{enumerate}
\item[1.] Choose a clone to evolve such that $j={\rm argmin}_i t^{(i)}$.
\item[2.] Compute $y_j = \lfloor Y(c_j) + \epsilon  \rfloor$, 
where $Y(c_j)=e^{\Delta t(c_j)\, \delta r_{s}(c_j)}$ and 
 $\epsilon \in \mathcal{U}[0,1]$. $\Delta t(c_j)$ is the time spent by the clone $j$ in the configuration $c_j$ since its last configuration change and $\delta r_{s}(c_j)$ is given by Eq.~(\ref{eq:deltaR}).
\item[3.] If $y_j=0$, eliminate this copy, and if $y_{j}>0$, make $y_{j}-1$ new copies of this clone.
\item[4.] Change their configurations (from $c_{j}$ to $c_{j}'$), with probabilities $W_{s}(c_{j} \to c_{j}')/r_{s}(c_{j})$. 
\item[5.] Update the waiting time of the copies to $t^{(j)} + \Delta t$ where $\Delta t$ is chosen from a exponential law of parameter $r_{s}(c'_{j})$.
\item[6.] Finally, for $y_j = 0$ choose a clone $k$, $k\neq j$ and copy it. Meanwhile, if $y_j>1$ erase $y_{j}-1$ clones \textbf{(constant population approach)}.
\end{enumerate}


\subsection[\qquad \qquad \quad Scaling Method]{Scaling Method}
\label{sec:AppSM}
The procedure which allows to extract the infinite-time infinite-$N_c$ limit of the CGF estimator (according to Refs.~\cite{partI,partII}) is summarized as follows:

\begin{enumerate}
\item[1.] For each $N_c \in \vec{N}_{c}$, determine the CGF estimator $\overline{ \Psi_{s}^{(N_{c})}(t) }$ (as in Eq.~(\ref{eq:PSI1})) up to a final simulation time $T$ .
\item[2.] Fit the obtained $\overline{ \Psi_{s}^{(N_{c})}(t)}$'s with Eq.~(\ref{eq:tScal1}): 
$f_{t}^{(N_{c})} = f_{\infty}^{(N_{c})} + b_{t}^{(N_{c})}t^{-1}$
and determine in each case $f_{\infty}^{(N_{c})}$.
\item[3.] Fit the extracted $f_{\infty}^{(N_{c})}$'s with Eq.~(\ref{eq:nScal1}): $f_{\infty}^{(N_c)} = f_{\infty}^{\infty} + b_{\infty}^{(N_c)}N_c^{-1}$  and determine the infinite-time infinite-$N_c$ limit of the CGF estimator $f_{\infty}^{\infty}$.
\end{enumerate}

\section*{References}

\bibliography{ArticleJstat}

\end{document}